# Time-resolved spectroscopy with entangled photons

Dmitry A. Kalashnikov[1], Elizaveta V. Melik-Gaykazyan[1,2], Alexey A. Kalachev[3], Ye Feng Yu[1], Arseniy I. Kuznetsov[1], and Leonid A. Krivitsky[1,^]

[1]*Data Storage Institute, Agency for Science, Technology and Research (A\*STAR), 138634 Singapore*
[2] *Faculty of Physics, M. V. Lomonosov Moscow State University, 119991 Moscow, Russia*
[3]*Zavoisky Physical-Technical Institute of the Russian Academy of Sciences, Kazan 420029, Russia*

[^]*Leonid-K@dsi.a-star.edu.sg*

*Interaction of light with media often occurs with a femtosecond response time. Its measurement by conventional techniques requires the use of femtosecond lasers and sophisticated time-gated optical detection[1-3]. Here we demonstrate that by exploiting quantum interference of entangled photons it is possible to measure the phase relaxation time of a media on the femtosecond time scale (down to 100 fs) using accessible continuous wave laser and single-photon counting. We insert the sample in the Hong-Ou-Mandel interferometer[4] and infer the phase relaxation time from the modification of the two-photon interference pattern. In addition to its simplicity and ease of use, the technique does not require compensation of group velocity dispersion[5-8] and does not induce photo-damage of the samples. This technique will be useful for characterization of ultrafast phase relaxation processes in material science, chemistry, and biology.*

Upon coherent light excitation of a media, its evolution is described by two processes: the population decay due to spontaneous emission and dephasing (or disorientation) of dipole moments due to their interaction with the surrounding medium. The latter process is determined by phase relaxation time $T_2$, also referred to as the coherence time[1]. Accurate measurement of $T_2$ is essential for characterization of a number of processes including, atomic collisions, molecular vibrations, studies of surface states and others[9-14].

For many processes, $T_2$ lies within the femtosecond time scale and is typically measured by the time-resolved spectroscopy[1-3]. This technique is universally applied to both optically thin and thick samples, as it allows accounting for an additional broadening of resonance absorption lines[15]. However, implementation of the time-resolved spectroscopy faces a number of practical challenges. First, it requires the use of a femtosecond laser system with pulse duration significantly less than the $T_2$. Second, signal detection with adequate temporal resolution requires the use of nonlinear wave-mixing processes. Moreover, femtosecond pulses have an inherently high peak power, and one has to be careful not to damage and/or modify the sample under study.

Here we report on a practical alternative to the conventional femtosecond time-resolved spectroscopy. It allows measurement of the coherence time $T_2$ with femtosecond time resolution without the need of a femtosecond laser and a sophisticated detection system. We exploit the unique properties of quantum entanglement, which have already gained momentum in addressing a variety of practical applications, including secure communication[16-18], metrology[19,20] and sensing[21]. We generate entangled photons via spontaneous parametric down conversion (SPDC)[22] and build the two-photon interference setup, known as the Hong, Ou and Mandel (HOM) interferometer. In the HOM interferometer, two indistinguishable photons interfere on a 50/50 beam splitter and then are detected by two single photon photodetectors[4]. Destructive interference of probability amplitudes results in the observation of a pronounced dip in the dependence of coincidences of photocounts on the optical delay referred to as the HOM dip. Earlier studies were carried out on the propagation of entangled photons through a dispersive media, which revealed the effect of cancellation of group velocity dispersion in the HOM interference [5-8]. Here, for the first time, we expand the applicability of the HOM interference for the study of the dynamic characteristics of a resonant medium. We introduce the sample in one arm of the interferometer and show that it leads to the modification of the shape of the HOM dip. The dip becomes asymmetric, elongated, and it demonstrates pronounced oscillations. Using our theoretical analysis we infer the phase relaxation time $T_2$ on the femtosecond time scale even though entangled states of light are generated using a continuous wave (cw) laser.

We consider the HOM interferometer with a resonant medium in one of the arms, described by a transfer function $H(\omega)$, and an optical delay line in another, see Fig.1a. We assume that the SPDC is produced by a cw-laser, in which case the biphoton field has a strong frequency correlation. Then the coincidence count rate between the two detectors is given by (see Methods):

$$P_c(\tau) = \frac{1}{4} \int d\nu |F(\nu)H(\omega_0 - \nu)|^2 + |F(\nu)H(\omega_0 + \nu)|^2 - 2\Re\{|F(\nu)|^2 H^*(\omega_0 - \nu)H(\omega_0 + \nu)e^{-i2\nu\tau}\}, \quad (1)$$

where $\omega_0$ is the central frequency of the biphoton field, $\nu$ is the frequency detuning, $F(\nu)$ is the biphoton field amplitude, $\tau$ is the time delay.

We assume that the resonant medium is a two-level system with a Lorentzian line shape

$$H(\omega_0 + \nu) = exp\left[-\frac{ib}{\nu - \Omega + i/T_2}\right], \quad (2)$$

where $b = \alpha L / 2T_2$, $\alpha L$ is the optical thickness ($\alpha$ is a Bouger coefficient and $L$ is the length of the media, and $\alpha L < 1$ for an optically thin sample), $T_2$ is the coherence time, and $\Omega \equiv \omega_{res} - \omega_0$, where $\omega_{res}$ is the resonant frequency.

Substituting Eq.2 in Eq.1 we can estimate the coherence time $T_2$ of the medium under study from the fitting parameters. To prove this concept we performed experiments with two different samples: an Nd:YAG crystal and an array of nanoparticles made of amorphous silicon (see Methods).

First, we measure the spectrum of the biphoton field and the transmission of the filter, see Fig.2a. From this data, we find the spectral profile of the biphoton field $F(\nu)$ (see Methods).

Then we measure the HOM dip without any sample, see Fig.2b. Our results show good quality of the entanglement with the uncorrected interference visibility of 92±0.3%. The experimental results in Fig.2b are fitted by Eq.1, assuming that in the absence of a sample *H=1*. The fit yields a coefficient of determination $R^2$=0.988. Subsequently, parameters of the biphoton field are used in the fitting of the experiments with the samples.

Then, we perform an experiment with the Nd:YAG crystal. We measure the transmission spectrum of the crystal using the spectrometer, and find that there are five absorption lines within the biphoton spectrum, see Fig.3a. From the obtained spectrum we determine the detuning of each line from the central wavelength of the biphoton field and the sample's optical thickness. Then, we measure the HOM dip with the Nd:YAG crystal in one of the arms of the interferometer. We find that the dip takes an asymmetric and elongated shape, see Fig.3b. This is attributed to a coherent resonant response of the medium[23]. Then we fit the interference pattern with Eqs.1-2 with the parameters defined from the spectrum and infer $T_2$ for each line. The best fit yields $R^2$=0.933 with the following values $T_2^I$=620±50 fs, $T_2^{II}$=660±50 fs, $T_2^{III}$=415±30 fs, $T_2^{IV}$=710±60 fs, $T_2^V$=215±20 fs, where roman numerals denote corresponding resonant lines. We highlight that strong absorption in the sample results in broadening and overlap of spectral lines. In this case, coherence times cannot be directly estimated from the spectral linewidths.

To ensure the validity of our method, we compare our results for the line IV ($\alpha L^{IV}$=7.2) with spectroscopic data obtained for an optically thin Nd:YAG sample with similar Nd concentration[24]. This line has the strongest absorption among others, and it is used for pumping of solid-state lasers. Based on the literature data, for an optically thin sample line IV has the FWHM of about 1 nm, which yields $T_2^{IV}{}_{thin\ sample}$≈700 fs[24]. This value is consistent with our measurements ($T_2^{IV}$=710±60 fs), thus ensuring the validity of our technique.

Next, we apply the developed methodology to the case when the optical thickness of the sample cannot be reduced without significant modification of its properties. We perform the experiment with an array of Silicon nanodiscs with diffractively coupled magnetic dipole resonances (see Methods). Following the procedure described above, we first measure the transmission profile of the sample, which possesses a single dip at 818 nm, see Fig.4a. We then measure the HOM dip, which has elongated and asymmetric form, see Fig.4b. We fit the obtained results using Eqs.1-2 with $T_2$ as a single fitting parameter. The fit yields $T_2$=130±15 fs with $R^2$=0.971. The measured $T_2$ is attributed to a dephasing time between the coupled magnetic dipole moments inside neighboring Si nanoparticles constituting the resonant mode. It is analogous to the dephasing time of surface plasmon polaritons in metal nanostructures and nanoarrays, which has been measured earlier using conventional ultrafast spectroscopy techniques[25-27]. Note, that for optically thin samples ($\alpha L < 1$), coherence time and width of the resonance are related as $T_2 = 1/\pi\Delta\nu_{1/2}$[1]. Considering a measured value of the FWHM of 12 nm from Fig.4a, we obtain $T_2$≈70 fs, which is almost two times smaller than our experimental result. This discrepancy occurs due to the optical thickness of the sample being larger than unity ($\alpha L = 4$), which broadens the linewidth and makes direct calculation of $T_2$ from the absorption spectra inadequate.

The temporal resolution of our technique is defined by the width of the spectrum of the biphoton field. In our experiment, it is 20 nm, which corresponds to a resolution of about 35 fs. It is on a par with existing high-performance femtosecond laser setups. With readily available methods for generation of broadband biphoton fields, it is feasible to achieve the temporal resolution down to a few femtoseconds or even less[28-30]. For example, SPDC generated in chirped crystals[28] yields the spectral width of 300 nm, which corresponds to the temporal resolution of 7 fs. Both, arbitrary shapes of resonant lines and spectral shape of the biphoton field, can be always accounted in numerical calculations see Eq.1.

In conclusion, we have demonstrated a new technique for measuring the phase relaxation time of a matter on the femtosecond time scale. Our approach utilizes the effect of the quantum two-photon interference of entangled photons and it allows excluding complex and expensive femtosecond laser setups. The technique allows the measurement of coherence times in optically thick samples, for which application of transmission spectroscopy is limited. Moreover, our approach does not suffer from the group velocity dispersion, which is a limiting factor in conventional methods. The technique operates at a single photon level and it can be useful for measurements of fragile biological, chemical and nanostructured samples. We believe that the technique will contribute to further development of ultrafast time-resolved spectroscopy in material science, biology, and chemistry.

## Methods

### Theory

We consider the HOM interferometer with a resonant medium in one of the arms, see Fig.1a. A biphoton field, produced via SPDC, can be represented as

$$|\psi\rangle = \iint d\omega_1 d\omega_2 F(\omega_1, \omega_2)|\omega_1\rangle|\omega_2\rangle = \iint d\omega_1 d\omega_2 F(\omega_1, \omega_2) a_1^\dagger(\omega_1) a_2^\dagger(\omega_2)|0\rangle, \quad (3)$$

where $F(\omega_1, \omega_2)$ denotes the biphoton field amplitude and $a_i^\dagger(\omega_i)$ denotes the creation operator at frequency $\omega_i$, $i=1,2$.

In one arm of the interferometer, we introduce an optical delay element with the transmission function $\phi(\omega)$ (in the absence of losses $|\phi(\omega)| = 1$). In another arm we introduce a resonant medium described by a transfer function $H(\omega)$. Then Eq.3 can be rewritten as:

$$|\psi\rangle = \iint d\omega_1 d\omega_2 F(\omega_1, \omega_2) a_1^\dagger(\omega_1) a_2^\dagger(\omega_2) \phi(\omega_1) H(\omega_2)|0\rangle. \quad (4)$$

After passing through the beamsplitter the state changes to

$$|\psi\rangle = \iint d\omega_1 d\omega_2 F(\omega_1, \omega_2) \left[ b_2^\dagger(\omega_1) b_1^\dagger(\omega_2) - b_1^\dagger(\omega_1) b_2^\dagger(\omega_2) + i b_1^\dagger(\omega_1) b_1^\dagger(\omega_2) + i b_2^\dagger(\omega_1) b_2^\dagger(\omega_2) \right] \phi(\omega_1) H(\omega_2)|0\rangle, \quad (5)$$

where $b_i^\dagger(\omega_i)$ is a creation operator at the output of the beamsplitter. The coincidence count rate between the two detectors is determined by the first two terms in Eq.5. After redefining the integration variables, the relevant part of the state can be written as follows:

$$|\psi_c\rangle = \tfrac{1}{2} \iint d\omega_1 d\omega_2 [F(\omega_1,\omega_2)\phi(\omega_1)H(\omega_2) - F(\omega_2,\omega_1)\phi(\omega_2)H(\omega_1)] |\omega_1,\omega_2\rangle. \tag{6}$$

We can now analyze the dependence of the coincidence count rate on the time delay, assuming that in a typical experiment the time window of the coincidence circuit (typically a few ns) is much larger compared to the coherence time of the field. In this case, the coincidence count rate is given by:

$$P_c = \iint d\omega_1 d\omega_2 |\langle \omega_1,\omega_2|\psi_c\rangle|^2. \tag{7}$$

Substituting $|\psi_c\rangle$ and taking into account that in the absence of losses $|\phi(\omega)| = 1$ we can rewrite Eq.7 in the following way

$$P_c = \tfrac{1}{4} \iint d\omega_1 d\omega_2 |F(\omega_1,\omega_2)H(\omega_2)|^2 + |F(\omega_2,\omega_1)H(\omega_1)|^2 - 2\Re\{F^*(\omega_1,\omega_2)F(\omega_2,\omega_1)H^*(\omega_2)H(\omega_1)\phi^*(\omega_1)\phi(\omega_2)\}, \tag{8}$$

where we assume that the time delay is introduced by an optical element without dispersion, and $\phi(\omega) = e^{i\omega t}$. For the SPDC pumped by a cw-laser the biphoton field has a strong frequency anticorrelation and the state described by Eq.3 can be represented as

$$|\psi\rangle = \int d\nu F(\nu)\, a_1^\dagger(\omega_0 - \nu) a_2^\dagger(\omega_0 + \nu)|0\rangle,$$

where $\omega_0$ is the central frequency of the biphoton field, and $\nu$ is detuning from the central frequency. Then, Eqs.4-8 can be rewritten in terms of $\omega_0$ and $\nu$, and the coincidence count rate is given by Eq.1.

**Spectral function of the biphoton field**

The spectral amplitude of biphoton field $F(\nu)$ consists of two components: the phase matching function of the nonlinear crystal $F_{bp}(\nu)$, which determines the width of SPDC spectrum, and the transfer function of the filter, placed in front of the detector $\Phi(\nu)$. This yields

$$F(\nu) = F_{bp}(\nu)\Phi(\nu). \tag{9}$$

In type-I frequency-degenerate SPCD, which is used in our experiment, the photons in a pair have the same polarization so that

$$F_{bp}(\nu) \propto L_c \chi^{(2)} E_{pump} \mathrm{sinc}\left(\frac{\nu^2 D'' L_c}{2}\right),$$

where $L_c$ is the length of the nonlinear crystal, $\chi^{(2)}$ is the nonlinear susceptibility, $E_{pump}$ is the pump field amplitude, $D'' = d^2k/d\nu^2$ is the group velocity dispersion in the crystal at the frequency of the SPDC field. The SPDC field is restricted by a filter with a trapezoidal shape $|\Phi(\nu)|^2$, where an imaginary part of $\Phi(\nu)$ is determined by the Hilbert transformation. The spectral function of the biphoton field is derived from Eq.9 using the measured SPDC spectrum and the transmission curve of the filter, see Fig.2a. The former has a full width on a half maximum $\Delta\lambda_{1/2}$=22 nm, and the latter is modeled by a trapezoidal function with a top width of 15.5 nm and side slopes of 3.3 nm.

**Samples**

The Nd:YAG crystal (Nd concentration 1%; 8 mm length; antireflection coated facets for 800 nm) has a strong absorption line at 808 nm and four satellite lines in the range 804-822 nm, which are within the spectral range of the biphoton field. There are five separate resonant absorption lines in the spectrum of Nd:YAG crystal within FWHM of the interference filter, see Fig.3a, where the roman numerals denote the corresponding line numbers. From the obtained spectrum we determine the detuning from the center of the biphoton field for each line: $\Omega^{I}$=7.5 nm, $\Omega^{II}$=3.1 nm, $\Omega^{III}$=-1.55 nm, $\Omega^{IV}$=-6 nm, $\Omega^{V}$=-9.9 nm, where the upper index denotes the line number. We estimate the corresponding optical thicknesses to be $\alpha L^{I}$=1.95, $\alpha L^{II}$=2.35, $\alpha L^{III}$=2.9, $\alpha L^{IV}$=7.2, $\alpha L^{V}$=3.6.

The second sample consists of Si nanodiscs with a diameter of 200 nm, disc height of 150 nm and pitch 550 nm on a quartz substrate (see inset at Fig.4a) covered by a polydimethylsiloxane (PDMS) layer for refractive index matching. The sample exhibits a strong narrow resonant dip in transmission at 818 nm. This dip corresponds to a mode inside the array excited due to the interaction of magnetic resonances in the single nanoparticles coupled through a diffraction order propagating along the array[31-35]. From the measured spectrum, we obtain the detuning from the central frequency of the biphoton field ($\Omega$=4.4 nm) and the optical thickness of the sample ($\alpha L$=4). It is important to note that since this sample consists of a single layer of resonant nanoparticles and its transmission is determined by the resonant interactions of the nanoparticles within the array, the optical thickness of the sample cannot be further reduced without losing its resonant properties. Thus this case corresponds to a situation when conventional transmission spectroscopy cannot be adequately applied for identification of $T_2$ while the proposed methodology can successfully accomplish this task.

**Experimental setup**

We produce photon pairs in a 0.8 mm long BBO crystal (Dayoptics) cut for type-I SPDC ($e \rightarrow oo$) and pumped by a CW-laser at 407 nm (PhoxX 405-60 Omicron), see Fig.1b. The pump laser power is set at 60 mW and the beam diameter is of 1.4 mm. Photons pairs are produced in a frequency degenerate non-collinear regime with emission angles $\theta = \pm 3$ deg with respect to the pump beam. The photons are coupled into two single mode fibers (SMF) with fiber paddles in both arms used for polarization control. The SPDC spectrum has a bandwidth of 22 nm, as it is defined by the phasematching conditions and coupling to SMFs[36]. Additionally, we use an auxiliary laser (at 814 nm) which facilitates alignment of the interferometer (not shown). The beams at the input of the HOM interferometer have the same polarization. The half wave plate (HWP1) set at 45 deg is placed in one of the arms, and the beams are recombined on a polarizing beamsplitter (PBS1). In one arm of the interferometer we introduce a delay line using a motorized translation stage (Owis) with a translation step of 0.5 μm. In another arm we place a telescopic 1:1 imaging system, consisting of two lenses with $f$=50 mm, which focuses the beam onto the sample (a spot size at the sample is 16 µm). A light from a halogen lamp (HL) is fed into the imaging system with a dichroic mirror (DM) and the image of the sample is captured by a CCD camera (Thorlabs). The beam combined at PBS1 passes through a half-wave plate (HWP2) set at 22.5 deg, which rotates the polarization by 45 deg and makes the photons indistinguishable. The photons are then split at a polarizing beamsplitter (PBS2) and coupled into two single mode fibers. We use an interference bandpass filter (IF; FF01-820/12-25, Semrock) tilted by 9 deg to tune its central transmission wavelength to 815 nm with FWHM=21 nm. In the setup there is a possibility either (1) to connect the output of the fibers to a home-build grating spectrometer (resolution of 0.2 nm) for the spectral measurements, or (2) directly to single photon avalanche photodetectors (D1, D2; SPCM-AQR-14FC, Perkin Emler) for HOM dip measurements. Signals from the detectors are sent to a coincidence counting scheme (Ortec, TAC 556) with a time window of 3 ns. Typical acquisition time for each sample was about 10-15 min, depending on the sample transmission. The obtained results are then fitted with the theory using Matlab and Origin software.

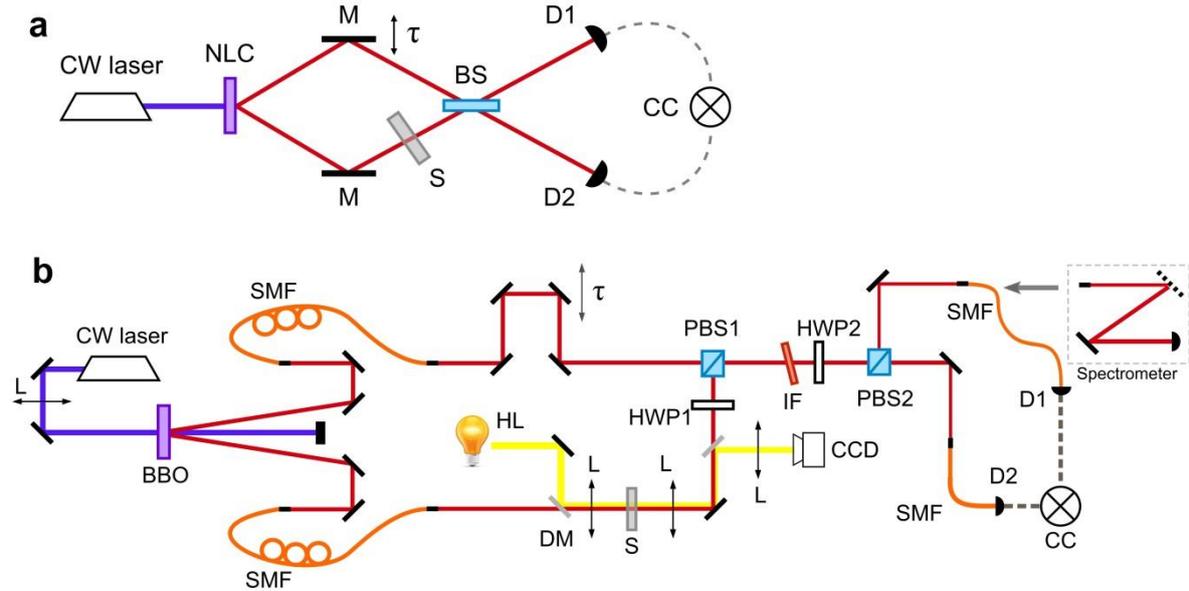

**Figure 1 | The experimental setup**. **(a)** Principal scheme. The pump from a CW-laser passes through a nonlinear crystal (NLC). The SPDC radiation is fed into the interferometer by a set of mirrors (M). Sample (S) is placed in one of the arms of the interferometer. The time delay τ is induced by a mirror on a translation stage. Photons interfere at a 50/50 beamsplitter (BS) and then are detected by avalanche photodetectors (D1 and D2). Signals from detectors are sent to a coincidence counting scheme (CC). **(b)** Experimental setup. A 407 nm cw-laser is focused by a lens (L) onto a BBO crystal cut for type-I SPDC. The SPDC radiation is fed into two single mode fibers (SMF) and then coupled into the interferometer. Sample (S) is placed in one of the arms of the interferometer. An imaging system consisting of two confocal lenses (L) allows observation of a sample on a CCD camera. Time delay τ is introduced by a motorized translation stage. The half-wave plate (HWP1) rotates the polarization at 90 deg and the two beams in the interferometer are combined on a polarization beamsplitter (PBS1). A half-wave plate (HWP2) rotates polarization at 45 deg, and the interference is observed at the output of the PBS2. In HOM interference the photons are detected directly by avalanche photodiodes (D1, D2) connected to a coincidence circuit (CC); for spectral measurements we use a grating spectrometer.

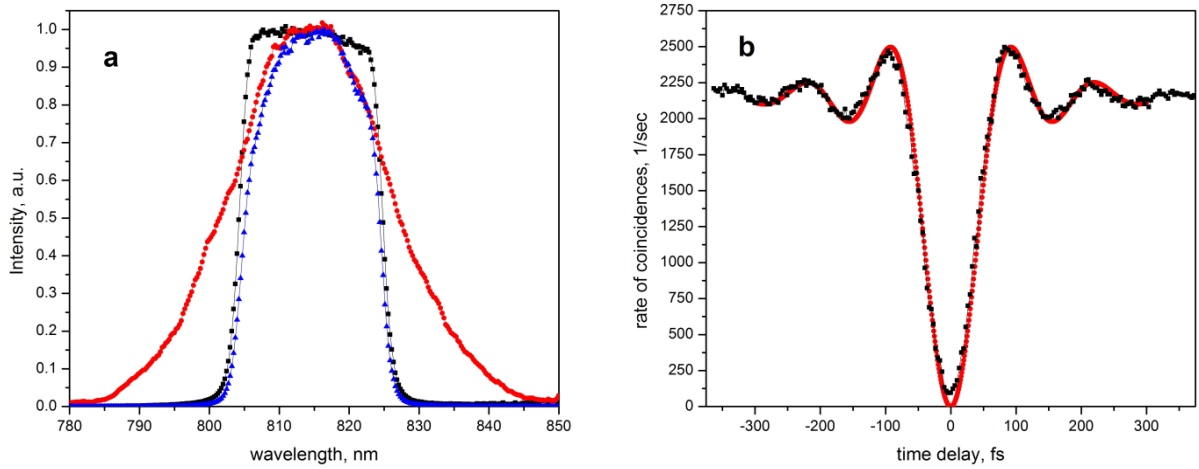

**Figure 2 | Characterization of the biphoton field (a)** Normalized transmission of the interference filter (black) and spectrum of SPDC (red). Their convolution (blue) determines the spectral shape of the biphoton field used in the experiment. **(b)** The HOM dip without samples. Experimental results (black) are fitted by Eq.1 (red) with measured parameters of the biphoton field.

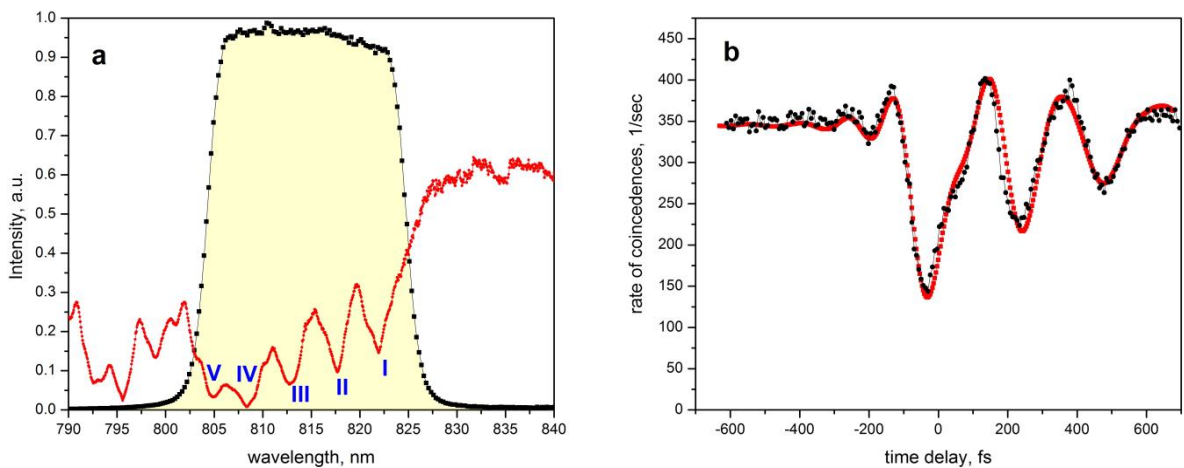

**Figure 3 | Results for the Nd:YAG crystal. (a)** The transmission spectrum of the 8 mm long Nd:YAG crystal (red), restricted by an interference filter (black dots and yellow shaded area). The intensity is normalized to the transmission by the filter. Blue roman numbers denote individual absorption lines. **(b)** The HOM dip with the Nd:YAG crystal in one arm of the interferometer. Experimental results (black) are fitted by Eqs.1-2 (red) with the fixed parameters of the biphoton field and the sample, and $T_2$ for each of the five lines being a fitting parameter.

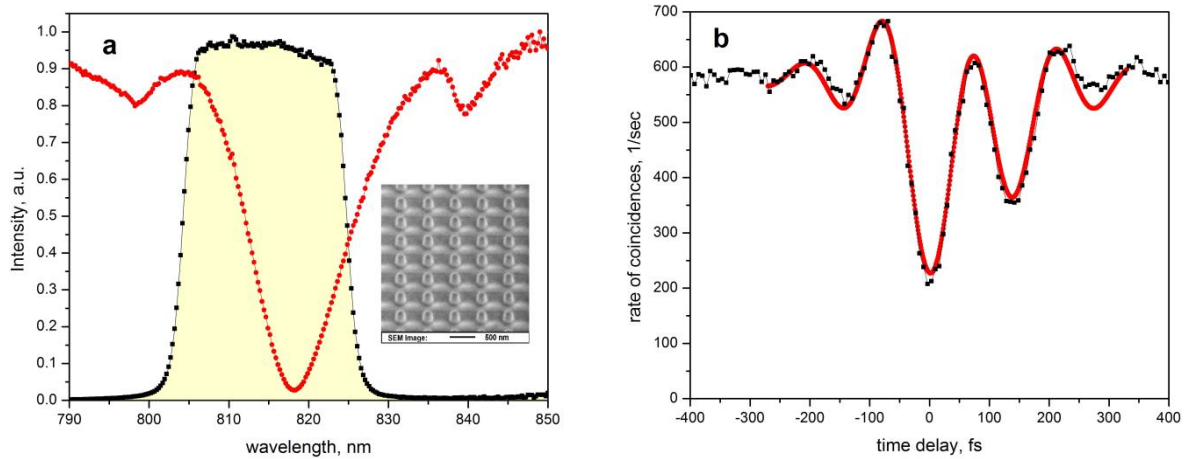

**Figure 4 | Results for the nanostructure. (a)** The transmission spectrum of the nanostructure (red), restricted by the interference filter (black dots and yellow shaded area). The intensity is normalized to the transmission of the filter. The inset shows a SEM image of the array of Si nanodiscs with the diameter of 200 nm, disc height of 150 nm and pitch of 550 nm. **(b)** The HOM dip obtained with the nanostructure in one arm of the interferometer. Experimental results (black) are fitted by Eqs.1-2 (red) with the measured parameters of the biphoton field and the sample. The $T_2$ is inferred from the fit as the only free parameter.

## References

1) Diels, J.-C. & Rudolph, W. *Ultrashort Laser Pulse Phenomena: Fundamentals, Techniques, and Applications on a Femtosecond Time Scale.* (Academic Press, San Diego, 2006).
2) Abramczyk, H. *Introduction to Laser Spectroscopy.* (Elseiver, Amsterdam, 2005).
3) Demtroder, W. *Laser Spectroscopy.* (Springer, Berlin, 2008).
4) Hong, C. K., Ou, Z. Y. & Mandel, L. Measurement of Subpicosecond Time Intervals between Two Photons by Interference. *Phys. Rev. Lett.* **59**, 2044-2046 (1987).
5) Franson, J. D. Nonlocal cancellation of dispersion. *Phys. Rev. A* **45,** 3126 (1992).
6) Steinberg A. M., Kwiat, P. G. & Chiao, R. Y. Dispersion cancellation in a measurement of the single-photon propagation velocity in glass. *Phys. Rev. Lett.* **68**, 2421 (1992).
7) Grice, W. P. & Walmsley, I. A. Spectral information and distinguishability in type-II down-conversion with a broadband pump. *Phys. Rev. A* **56**, 1627 (1997).
8) Okamoto, R., Takeuchi, S. & Sasaki, K. Tailoring two-photon interference with phase dispersion. *Phys. Rev. A* **74**, 011801 (2006).
9) Di Bartolo, B. & Forte, O. *Frontiers of Optical Spectroscopy: Investigating Extreme Physical Conditions with Advanced Optical Techniques.* (Kluwer Academic Publishers, Dordrecht, 2005).
10) Shumay, I. L. *et al.* Lifetimes of image-potential states on Cu(100) and Ag(100) measured by femtosecond time-resolved two-photon photoemission. *Phys. Rev. B* **58**, 13974 (1998).

11) Stolow, A., Bragg, A. E. & Neumark, D. M. Femtosecond Time-Resolved Photoelectron Spectroscopy. *Chem. Rev.* **104**, 1719-1757 (2004).
12) Zewail, A. H. Femtochemistry: Atomic-Scale Dynamics of the Chemical Bond. *J. Phys. Chem. A* **104**, 5660-5694 (2000).
13) Gruebele, M. & Zewail, A. H. Femtosecond wave packet spectroscopy: Coherence, the potential, and structural determination. *J. Chem. Phys.* **98**, 883 (1993).
14) Holt, N. E., Kennis, J. T. M., Dall'Osto, L., Bassi, R. & Fleming, G. R. Carotenoid to chlorophyll energy transfer in light harvesting complex II from *Arabidopsis thaliana* probed by femtosecond fluorescence upconversion. *Chem. Phys. Lett.* **379**, 305-313 (2003).
15) Exton, R. J. Widths of Optically Thick Lines. *J. Quant. Spectrosc. Radiat. Transfer.* **11**, 1377-1383 (1971).
16) Ladd, T. D., Jelezko, F., Laflamme, R., Nakamura, Y., Monroe, C. & O'Brien, J. L. Quantum computers. *Nature* **464**, 45-53 (2010).
17) Gisin, N., Ribordy, G, Tittel, W. & Zbinden, H. Quantum Cryptography. *Rev. Mod. Phys.* **74**, 145-195 (2002).
18) Gisin, N. & Thew, R. Quantum communication. *Nature Photon.* **1**, 165-171 (2007).
19) Giovannetti, V., Lloyd, S. & Maccone, L. Advances in quantum metrology. *Nature Photon.* **5**, 222-229 (2013).
20) Georgiades, N. P., Polzik, E. S., Edamatsu, K., Kimble, H. J. & Parkins, A. S. Nonclassical Excitation for Atoms in a Squeezed Vacuum. *Phys. Rev. Lett.* **75**, 3426 (1995).
21) Kalashnikov, D. A., Paterova, A. V., Kulik, S. P. & Krivitsky, L. A. Infrared Spectroscopy with Visible Light. *Nature Photon.* **10**, 98-101 (2016).
22) Klyshko, D. N. *Photon and Nonlinear Optics*. (Gordon and Breach Science, New York, 1988).
23) Burnham, D. C. & Chiao, R. Y. Coherent Resonance Fluorescence Excited by Short Light Pulses. *Phys. Rev.* **188**, 667-675 (1969).
24) Lu, J. *et al.* Optical properties and highly efficient laser oscillations of Nd:YAG ceramics. *Appl. Phys. B* **71**, 469-473 (2000).
25) Busch K., Lolkes, S., Wehrspohn, R. B. & Foll, H. *Photonic Crystals: Advances in Design, Fabrication, and Characterization.* (Wiley, Weinheim, 2004).
26) Zentgraf, T., Christ, A., Kuhl, J. & Giessen, H. Tailoring the Ultrafast Dephasing of Quasiparticles in Metallic Photonic Crystals. *Phys. Rev. Lett.* **93**, 243901 (2004).
27) Anderson, A., Deryckx, K. S., Xu, X. G., G. Steinmeyer, G. & Raschke, M. Few-Femtosecond Plasmon Dephasing of a Single Metallic Nanostructure from Optical Response Function Reconstruction by Interferometric Frequency Resolved Optical Gating. *Nano Lett.* **10**, 2519-2524 (2010).
28) Nasr, M. B. *et al.* Ultrabroadband Biphotons Generated via Chirped Quasi-Phase-Matched Optical Parametric Down-Conversion. *Phys. Rev. Lett.* **100**, 183601 (2008).
29) Katamadze, K. G. & Kulik, S. P. Control of the spectrum of the biphoton field. *JETP* **112**, 20 (2011).

30) Okano, M., Okamoto, R., Tanaka, A., Subashchandran, S. & Takeuchi, S. Generation of broadband spontaneous parametric fluorescence using multiple bulk nonlinear crystals. *Opt. Express* **20**, 13977 (2012).
31) Evlyukhin, A. B., Reinhardt, C., Seidel, A., Luk'yanchuk, B. S. & Chichkov, B. N. Optical response features of Si-nanoparticle arrays. *Phys. Rev. B* **82,** 045404 (2010).
32) Garcia-Etxarri, A., Gomez-Medina, R., Froufe-Perez, L. S., Lopez, C., Chantada, L., Scheffold, F., Aizpurua, J., Nieto-Vesperinas, M. & Sáenz, J. J. Strong magnetic response of submicron silicon particles in the infrared. *Opt. Express* **19**, 4815–4826 (2011).
33) Kuznetsov, A. I., Miroshnichenko, A. E., Fu, Y. H., Zhang, J. B., & Luk'yanchuk, B. Magnetic Light. *Sci. Rep.* **2**, 492 (2012).
34) Evlyukhin, A. B., Novikov, S. M., Zywietz, U., Eriksen, R. L., Reinhardt, C., Bozhevolnyi, S. I. & Chichkov, B. N. Demonstration of magnetic dipole resonances of dielectric nanospheres in the visible region. *Nano Lett.* **12**, 3749–3755 (2012).
35) Kuznetsov, A. I., Miroshnichenko, A. E., Brongersma, M. L., Kivshar, Y. S. & Luk'yanchuk, B. "Optically resonant dielectric nanostructures", Science (2016), *in press*.
36) Kurtsiefer, Ch., Oberparleiter, M. & Weinfurter, H. High efficiency entangled photon pair collection in type II parametric fluorescence. *Phys. Rev. A* **64**, 010102(R) (2001).